\documentclass[preprint,pre,aps,superscriptaddress,floats,showpacs]{revtex4}
\usepackage{graphicx}
\usepackage{amsmath}
\usepackage{amsfonts}
\usepackage{amssymb}
\usepackage{epsfig}
\usepackage{color}
\usepackage{bm}

\begin{document}

\title{Laser-triggered THz emission from near critical density targets}

\author{V.~Yu.~Bychenkov}
\affiliation{P.~N.~Lebedev Physics Institute,
    Russian Academy of Science, Leninskii Prospect 53, Moscow 119991,
    Russia}
\affiliation{Center for Fundamental and Applied Research,
Dukhov Research Institute, Moscow 127055, Russia}
\author{A.~V.~Brantov}
\affiliation{P.~N.~Lebedev Physics Institute,
Russian Academy of Science, Leninskii Prospect 53, Moscow 119991,
Russia}
\affiliation{Center for Fundamental and Applied Research,
    Dukhov Research Institute, Moscow 127055, Russia}
\author{M.~G.~Lobok}
\affiliation{P.~N.~Lebedev Physics Institute,
Russian Academy of Science, Leninskii Prospect 53, Moscow 119991,
Russia}
\affiliation{Center for Fundamental and Applied Research,
    Dukhov Research Institute, Moscow 127055, Russia}
\email[Corresponding author: ]{mglobok@me.com}
\author{A.~S.~Kuratov}
\affiliation{P.~N.~Lebedev Physics Institute,
Russian Academy of Science, Leninskii Prospect 53, Moscow 119991,
Russia}
\affiliation{Center for Fundamental and Applied Research,
    Dukhov Research Institute, Moscow 127055, Russia}

\begin{abstract} Femtosecond laser pulse propagation in a relativistic self-trapping regime (RST) in a near-critical density plasma makes it possible to maximize the total charge of the accelerating electrons and laser-to-electrons conversion rate, that can be used to provide a large amount of the THz range coherent transition radiation. The three-dimensional particle-in-cell simulations demonstrate how such transition radiation generates when electrons escape into vacuum either from the low-density target itself, or after passing through a thin foil located at the target end. Advantage of the RST regime for generation of THz pulses is clearly demonstrated as compared to laser irradiation of such a standard target as a foil with preplasma on its front side. Simulation performed has shown that for the optimized laser-target matching a 2-J femtosecond laser pulse is able to produce quasi-unipolar Thz pulses with energy exceeding 100 mJ.
\end{abstract}


\maketitle

\section{Introduction}\label{sec1}

The sources of secondary radiation based on laser-generated electrons are seen as breakthrough for achieving a record brightness \cite{albert16}. For example, bunches of high-energy electrons produced by laser-wakefield accelerators (LWFA) or direct laser acceleration (DLA) mechanism are the sources of X-ray and gamma radiation suitable for many applications \cite{albert16,Rosmej22}. Also, when the LWFA or DLA electrons traverse the plasma-vacuum interface a broadband THz or far/mid-infrared radiation emits \cite{Leemans03,Schroeder04,Plateau09,Yang18,Dechard18,Pak23,Yang22, Gorlova,Gorlova24}. Energy of coherent transition radiation (CTR) pulse is proportional to the square of the electron bunch charge and increases with the electrons energy, that motivates for finding proper laser-target desings providing generation of ultra-relativistic electrons with the maximum possible total charge. However, typical LWFA gives relatively small charge of ultra-relativistic electrons (usually $\lesssim1$ nC) imposing undesirable limitations on practical applications of laser-triggered terahertz pulses, which are expecting to be the most intense over all known till now THz sources. This could make it possible not only considerably expand the widely discussed THz applications for various non-destructive diagnostics, but also develop the THz-driven strong-field physics, in particular, to manipulate of matter by changing its properties~\cite{salen}, especially, making use of a sub-cycle waveform of the THz-pulse \cite{kuratov22}.

In the first proof-of-principal LWFA experiment with a rather dense plasma, $\gtrsim 10^{19}$ cm$^{-3}$, and the laser pulse having a FWHM duration $\gtrsim$ 50 fs and a peak power $\lesssim 10$ TW \cite{Leemans03} the broadband terahertz pulse energy was limited by the narrow transverse dimension of the plasma to 3-5 nJ within a 30-mrad collection angle. Broadband terahertz pulse generation was concluded to be consistent with CTR associated with a longitudinal electron momentum distribution that was well described by a Boltzmann distribution with a temperature of a few MeV. Moreover, pulse energy was observed to scale quadratically with bunch charge ($Q_b$) in the range $0.02\, {\rm nC}\lesssim Q_b\lesssim 1.5\, {\rm nC}$, that also is consistent with CTR. Corresponding modeling indicated that this broadband source produced about 0.3 $\mu$J THz energy within a 100 mrad angle. Based on this, it was extrapolated that increase of the transverse plasma size and the mean energy of the electron beam to the few tens of MeV could further increase the pulse energy to the few 100 mJ level \cite{Leemans03,Schroeder04}. However, even now the LWFA generation of THz emission is still far from such yields, while subsequent experimental optimization of THz radiation generation from a laser wakefield accelerator performed in Ref. \cite{Plateau09} have led to production of a THz pulse with an energy of 5 $\mu$J corresponding to a conversion rate of $10^{-5}$.

Numerical simulations using the particle-in-cell (PIC) code for the wakefield-initiated CTR predicts a conversion efficiency to THz emission at the level of $10^{-3}$ \cite{Yang18,Dechard18}. A similar laser-to-terahertz conversion efficiency, up to $\sim 0.15$\%, has been demonstrated in the recent experiment with a 100 TW femtosecond laser pulse irradiating a nitrogen target \cite{Pak23}. There the THz signal of the 4 mJ energy correlated with electron bunch charge rather than with electron maximum energy and increased with electron plasma density. Based on the PIC simulations \cite{AIP_Advances_2024} the authors arrived to a surprising conclusion regarding the THz generation mechanism, explaining it by a group of emitting electrons which quickly accelerate by the ponderomotive force and plasma wakefield within a plasma rather than by CTR at a plasma-vacuum interface. Nevertheless, these results are still not competitive to those for more efficient THz generation from two-color laser filamentation in gases, where a conversion rate has been measured as high as $\gtrsim 2$ \% \cite{Koulouklidis} corresponding to the THz pulse energy of $\simeq 0.2$ mJ. Even somewhat higher conversion rate to THz radiation is achievable through optical rectification in organic crystals. For example, the highest to date laser-THz conversion rate of 3\%  has been demonstrated with electro-optic DSTMS crystals \cite{vicario}. However, such optical rectification method is not yet able to overcome the THz power upper limit of the order of 1 mJ unlike laser-plasma interaction mechanism of THz generation, which occurs without such kind limitation. On the other hand, generation of multi-mJ THz radiation in laser-plasma interaction has still not been accompanied by a breakthrough in the laser-to-THz conversion rate increase. For example, the most powerful THz pulse due to CTR with the energy of up to 200 mJ has been registered in the experiment performed at the Vulcan laser facility in the interaction of a 70 J picosecond pulse with a solid target \cite{Liao}. At the same time, the found conversion efficiency, $\sim 0.2$\%, was quite modest although the estimated charge of the escaping electron bunch was quite high, $Q_b \simeq 300$ nC. Indeed, in regards to a sufficiently long ps laser pulse, it is worth considering that an CTR energy actually scales as the product of the electron bunch charge and its linear charge rather than the square of $Q_b$. Accordingly, for the given electron bunch charge the THz yield from the femtosecond pulse dominates over that from the picosecond one.

Similar to the wakefield initiation of CTR the DLA electron acceleration mechanism is able to provide the same laser-to-THz conversion efficiency. This has been demonstrated in Ref. \cite{Gorlova}, where the conversion efficiency $\sim 0.2$\% was reached using a 1 TW, 40 fs laser pulse irradiating a tape with artificially produced preplasma by a synchronized nanosecond pulse \cite{Gorlova}. Such a fairly high efficiency was due to DLA electrons producing in near-critical plasma and propagating through the target \cite{Gorlova24}. The reported enhanced THz emission from a two-layer target with a near-critical density first layer in accordance with the results of the 2D PIC simulation can also be associated with DLA electrons \cite {Yang22}.

Given the above, it is now motivated and important to maximize the CTR THz radiation yield by search of most effective way to increase the number of electrons escaping the target irradiated by a fs laser pulse. Based on the similar example with improving performance of a laser betatron X-ray source from the near-critical density plasma \cite{Vais} the same can be expected for THz emission by CTR when using a relativistic self-trapping (RST) regime in the form of "laser bullet" \cite{lobok18}. When making conditions for the implementation of such a regime of laser pulse propagation in a transparent but sufficiently dense plasma it has already been demonstrated how the best efficiency of electron acceleration in terms of the highest total charge of high-energy electrons and the highest laser-to-electrons conversion rate can be achieved. The universal way is a proper matching of laser hot spot size and electron plasma density to the laser pulse energy and duration. For example, following this way it can be provided with a 10-30 fs laser pulse a record number of accelerated electrons with energy in excess of 30 MeV with only $\sim 2$ J of laser energy. The RST mode physics details and corresponding 3D PIC simulations can be found in Refs. \cite{bychenkov19,bychenkov,bychenkov21}. Below, we study the CTR of electrons accelerated in the RST mode of laser pulse propagation in a plasma with a near critical electron density.

Another important issue that deserves attention is the waveform of the generated CTR THz signal. Despite extensive experimental evidence from experiments (e.g. \cite{Gao_OL_2008,Lei-iScience-2022,Gorlova}) and numerical simulations (e.g. \cite{Ding_APL_2013,Gorlova24}) about quasi-unipolar (half-cycle) CTR THz waveform, there was still no consensus on this issue and mainly because of lack of solid theoretical base. We have recently developed the analytical CTR theory for electrons escaping from a overdense target into a vacuum \cite{kuratov22}, which demonstrates the THz wave half-cycle pulse propagation from a target surface in a near-field zone. Here we study the spatiotemporal field distribution of the terahertz pulse generated through CTR of the electrons accelerated in the near critical density plasma by the femtosecond laser pulse propagating in the RST regime.

This paper is organized as follows. Section \ref{sec2} summarizes rough estimates for CTR of electrons accelerated in the RST mode of laser pulse propagation in a transparent plasma and during laser interaction with overdense foil. Simulation setups for low-density and overdense targets are described in Sec. \ref{sec3}. The PIC simulation results showing energy, space-time, angular and spectral characteristics of the generated THz pulse by electrons accelerated in the RST regime are presented in Sec. \ref{sec4}. Section \ref{sec5} compares these results with those for the THz CTR from a thin foil. We conclude with a summary in Sec. \ref{sec6}.

\section{Introductory remarks}\label{sec2}

Let us recall that the implementation of the RST mode of a pulse propagation in a plasma requires a matching of the formed plasma cavity diameter $D$ to the trapped laser light intensity $I = 1.37 (a_0/\lambda{\rm[\mu m]})^2\times 10^{18} $ W/cm$^2$ and the electron plasma density, $n_e$, so that laser beam diffraction divergence is balanced by relativistic nonlinearity. For the ultrarelativistic intensities ($a_0\gg 1$), this requirement confirmed by the theory \cite{bychenkov}
corresponds also to the upper limit of the cavity transversal size, which still does not allow the development of the laser beam filamentation,
\begin{equation} \label{rst1}
D \simeq \lambda_p= \lambda \frac{\omega}{\omega_p} \sqrt
{\gamma_e}\,,  \quad \gamma_e \gg 1\,,
\end{equation}
 where $\lambda_p$ is the relativistic plasma wavelength, $\lambda$ is the laser wavelength, $\omega_p$ is the electron plasma frequency, $a_0 = e E/m \omega c \gg 1$ is the standard dimensionless laser field amplitude, $\omega$ is the laser field frequency and $\gamma_e = \sqrt{1+a_0^2/2} \simeq a_0/\sqrt{2}$ is the electron relativistic factor. The relation Eq. (\ref{rst1}) can be rewritten in the form
 \begin{equation} \label{rst}
 D \simeq \lambda \,\sqrt{\frac{n_c}{n_e}\frac{a_0}{\sqrt{2}}}\,, \quad a_0 \gg 1\,,
\end{equation}
 where $n_c$ - is the laser critical density.

 Note that execution of the condition Eq. (\ref{rst}) requires the laser focal spot size $D_L$ at the plasma entrance to be somewhat less than $D$, namely the cavity diameter slowly increases over time from $D_L$ approximately up to $D\sim 2 D_L$, as shown in Ref.  \cite{lobok18}. The RST mode may exist in two typical forms: (1) "bubble", when the pulse length $c\tau$ is significantly less than the cavity length $L$, which in turn is equal to the cavity transverse size $D$ (spherical cavity, bubble), $c\tau \ll L \simeq D$, \cite{Pukhov_2002} and (2) "laser bullet", when a cavity is entirely filled by the laser field, $c\tau \simeq L$ \cite{lobok18,bychenkov19}. Certainly, the intermediate cases, $c\tau < L$, are also acceptable.
 \begin{figure}[!ht]
 	\includegraphics[width=0.8 \textwidth]{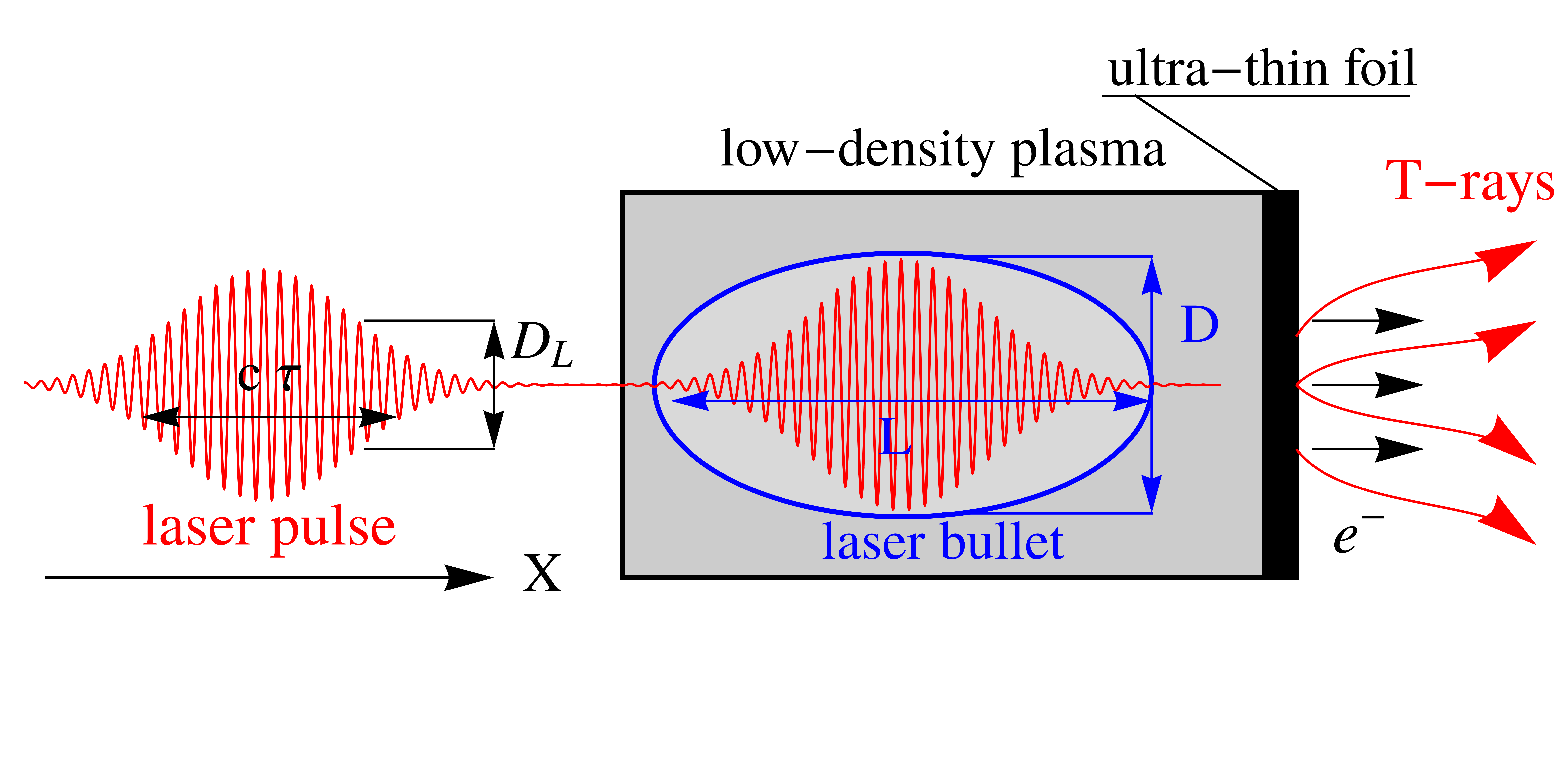}
 	\caption{Schematic illustration of the THz-generation scenario through CTR by the electrons based on the "laser bullet" RST regime.
 	} \label{fig0}
 \end{figure}
To avoid longitudinal self-modulation instability of a "laser bullet" its length, $c\tau$, similar to Eq. (\ref{rst}), should be limited, $c\tau\simeq \lambda_p \simeq D$. Thus, a long "laser bullets", $c\tau >D$, are prohibited and one arrives to the matching conditions for the "laser bullet" of a spherical shape
 \begin{equation} \label{rst2}
	c\tau \simeq D \simeq \lambda \,\sqrt{\frac{n_c}{n_e}\frac{a_0}{\sqrt{2}}}\,, \quad a_0 \gg 1\,.
\end{equation}
In the simulations presented below, we consider the "laser bullet" RST regime following the matching conditions Eq. (\ref{rst2}). This regime slowly evolves to the "bubble" regime at the final stage as laser pulse depletes during propagation. Schematic illustration of the THz-generation CTR scenario by the electrons accelerated in the "laser bullet" is presented in Fig. \ref{fig0}.

Emitted CTR energy by relativistic electron bunch escaping from a target into a vacuum is mainly depends on the bunch length and its total charge in accordance with the estimate
\begin{equation}\label{W}
	W_{THz} = \frac{Q_b^2}{\pi L_b} \left \{ 3 \ln \left(\frac{\sqrt{2}{\cal E}_b L_b}{m c^2 D_b}  \right) -1 \right \} \,,
\end{equation}
where ${\cal E}_b$ is its characteristic energy, and $L_b$ and $D_b$ are the bunch length and width, respectively. The estimation Eq. (\ref{W}) is somewhat distinguished (by a numerical factor in front of $Q^2_b$ and by the $L_b/D_b$ ratio under the logarithm) from the qualitative ones based on the single particle approach \cite{Schroeder04,Pak23}. The details of such quantitative adjustment can be found in Refs. \cite{kuratov16,kuratov22}.

Important issue naturally raises regarding the value of the charge $Q_b$ since a sharp dependence on it of the emitted energy Eq. (\ref{W}). To leave a target, electron must have enough energy to overcome the attraction potential, $\phi \simeq 2Q/D$, of a positively charged cavity (with charge $Q$) reaching the rear boundary of the target. In accordance with Eq. (\ref{rst2}), the threshold electron energy, $\epsilon_{min}$, required to leave the target, that comes from the balance $\epsilon_{min}\simeq 2eQ/D$ reads
\begin{equation}\label{epsmin}
	\epsilon_{min} \simeq \frac{\pi^2}{3 \sqrt{2}}a_0mc^2 \,.
\end{equation}
For example, the estimate Eq. (\ref{epsmin}) gives $\epsilon_{min} \simeq$ 30 MeV for $a_0=24$. To ensure the adequacy of both the above estimates and the PIC simulation results one may find $Q_b$ from the numerical electron spectrum, $dN/d\epsilon$,
\begin{equation}\label{Qf}
	Q_b = e\int_{\epsilon_{min}}^{\infty}\!\!\!d\epsilon \, \frac{dN}{d\epsilon} \,,
\end{equation}
by using Eq. (\ref{epsmin}) and compare corresponding values of the THz energy yields.

The relation for the emitted terahertz energy Eq. (\ref{W}) has been derived for the case of high electrical conductivity of the target, i.e. for the medium with a dielectric permittivity $|\epsilon| \gg 1$. This is well suited up to the CTR cutoff frequency $\omega \simeq c/L_b$ for the considering here densities $n_e \gtrsim 0.1$ and multy-micrometer electron beam lengths, that for $\omega_{pe} \gtrsim 6 \times 10^{14}$ c$^{-1}$ and $L_e\sim 10\lambda$ corresponds to $|\epsilon| =| 1 -\omega_{pe}^2/\omega^2 |\approx \omega_{pe}^2/\omega^2 \gtrsim 4(L_e{\rm[\mu m]})^2\gg 1$.

Below we not only study the THz CTR generation from the "laser bullet" RST regime but also compare it with more popular scheme for the THz CTR production from thin laser-irradiated foil. It is well known that for most effective generation of high-energy electrons providing high THz CTR yield the using proper adjusted preplasma size is required \cite{Sentoku}. Such preplasma assumed in the estimates below may heat significant amount of electrons, that leads to a few tens percent laser light energy conversion to hot electrons energy.

To leave a thin overdense target the laser heated electrons must have high enough energy to overcome the back side sheath field potential. Sheath field effectively slows down accelerated electrons at the target back side in the double layer with high decelerating Coulomb field at the distance from the target $x\lesssim D_b$. Only those electrons that have sufficient energy to overcome the distance $D_b$, at which the Coulomb attraction starts to drop, leave the target. One can make something like similar comparison as above, but for the CTR-generating electrons accelerated from laser - thin foil interaction. In this case, the minimum energy of escaping electrons can be estimated by the balance $\epsilon_{min} \simeq -e \phi (x \simeq D_b)$.
For the widely assumed Boltzmann distribution of laser heated electrons with the temperature $T_h$ and density $n_h$, we use
well-known solution for the sheath potential $e \phi(x) = -2 T_h \ln (\sqrt{\mbox{e}} + x/(\sqrt{2} \lambda_{Dh}))$, where $\lambda_{Dh} = \sqrt{T_h/4 \pi e^2 n_h}$ is the hot electron Debay radius. We also use another commonly accepted ponderomotive scaling for the hot electron temperature $T_h \simeq m c^2 \sqrt{1+a_0^2/2} \simeq m c^2 a_0/\sqrt{2}$. Then, from the balance, $\eta I_L \simeq c n_h T_h$, of the laser and particle energy fluxes, where $\eta$ is the laser light conversion (absorption) coefficient and $I_L$ is the laser pulse intensity, one may estimate
\begin{equation}
\epsilon_{min} \simeq \sqrt{2} m c^2 a_0 \ln \left(\sqrt{\mbox{e}} + \pi\sqrt{2}\,\frac{D_b\sqrt{\eta}}{\lambda} \right) \,.
\end{equation}
For example, $\epsilon_{min} = 30$ MeV for $a_0=24$, $D_b =4 \lambda$ and $\eta =0.2$.

Note that the bulk of the laser-heated electrons in the double layer make small contribution to the THz-pulse generation since compensation of the electron current by return electrons and radiation field suppression \cite{Denoual23} due to small Debye layer thickness, $\lambda_{Dh} \ll c \tau$.

\section{Simulation setup}\label{sec3}

Below, the results of the performed 3D PIC simulations study of the THz generation from laser-plasma interaction for both transparent near critical density target and thin overdense plasma foil with a preplasma are presented. The same pulse energy of 2.2 J was used for these two target designs.

As for the first case here, the laser-plasma parameters were chosen to match the RST mode. Laser pulse propagating in the X-direction and polarized  along Z-axis was focused on the front side of a homogeneous plasma target consisting of electrons and He-ions. Two different pulse durations have been considered, 30 fs laser pulse with $a_0 = 24$ focused in $D_L= 2.8 \lambda$ (FWHM) spot and 10 fs laser pulse with $a_0 =41.57$ focused in  $2.8 \lambda$ spot. The optimal plasma densities relevant to Eq. (\ref{rst}) were chosen $0.1 n_c$ and $0.15 n_c$ for the 30 fs and 10 fs laser pulses, correspondingly. In accordance with the depletion length estimates \cite{bychenkov19} corresponding plasma lengths were $240 \lambda$ and $100 \lambda$. The simulations were performed by using the moving window technique with the spatial grid steps $0.02 \lambda \times 0.1 \lambda \times 0.1 \lambda$ in a simulation window $X \times Y \times Z = 78 \lambda \times 118 \lambda \times 118 \lambda$. Simulation window is stopped when its left boundary approaches plasma-vacuum interface. For each particle species 2 numerical particles were used per cell. A single typical simulation run required on average about 100 hours of computation on computing cluster with 200 CPU cores.

For comparison reason,  simulations with additional overdense thin plasma foil ($1\,\mu$m thickness) at the plasma back side have been performed. This make it possible to understand the effect of increase of target electric conductivity on the CTR radiated field and cut-off a residual laser field leaking from the target back side into the vacuum.

For the second case, the pulse with the parameters $\tau$=30 fs, $a_0=24$, and $D_L= 3.3 \lambda$ has been used to estimate the THz yield when it irradiates a thin aluminum foil with a preplasma. As a main overdense target we used plasma layer with the thickness of $4 \lambda$, which consist on Al$^{+11}$ ions and electrons with electron density of $200 n_c$. Since the most energetic electrons responsible for THz production are accelerated within the preplasma layer, the used high density of the main foil does not affect a THz waves generation. We used preplasma with a thickness of $45 \lambda$ having a bi-exponential density profiles with two characteristic density gradients associated with the near-critical region ($0.4 \lambda$) and the subcritical extended plasma corona ($14 \lambda$) \cite{rakitina}. The simulation box was $75 \lambda \times 30 \lambda \times 30 \lambda$ with the spatial grid steps $0.02 \lambda \times 0.05 \lambda \times 0.05 \lambda$. Eight particles per cell has been used for both ions and electrons. To estimate effect of the preplasma scale length on electron acceleration a short preplasma with sharp density profile was also considered for comparison.

\section{PIC simulation of the THz CTR by electrons accelerated in the RST regime}\label{sec4}

\begin{figure}[!ht]
\includegraphics[width=0.7 \textwidth]{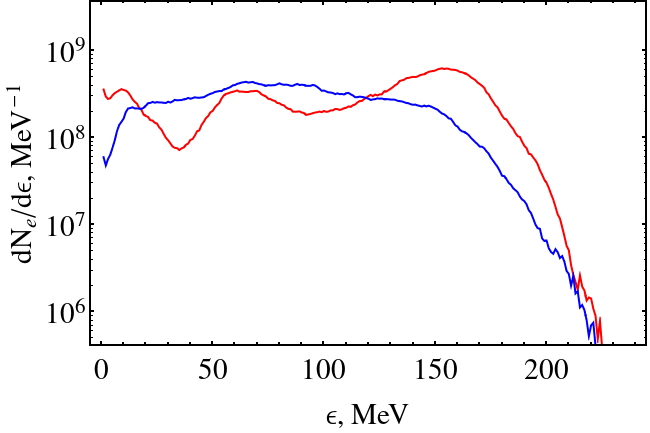}
\caption{Electron spectra generated in the RST regime by 30 fs (blue curve) and 10 fs (red curve) laser pulses.} \label{fig2}
\end{figure}

In accordance with the previous studies \cite{lobok18,bychenkov19}, very efficient electron acceleration in the RST regime has been reproduced in the performed simulations. For example, in the case of 10 fs laser pulse, the 9 nC electron bunch with the energies $>$ 1 MeV has been generated. The total $\sim 1$  J electron beam energy corresponds to the conversion efficiency of 46\%.  It is more efficient than for the 30 fs laser pulse, which produces electron beam with the charge 7.7 nC, total energy of 0.7 J and conversion rate $\sim 32$ \%. The latter clearly demonstrates an enhancement of electron efficiency production with pulse compression, e.g. by the CafCA method \cite{Khazanov19}. The electron spectra are shown in  Fig. \ref{fig2} for the instant when half of the electron beam has emerged from the target.

\begin{figure}[!ht]
\includegraphics[width=0.4\textwidth]{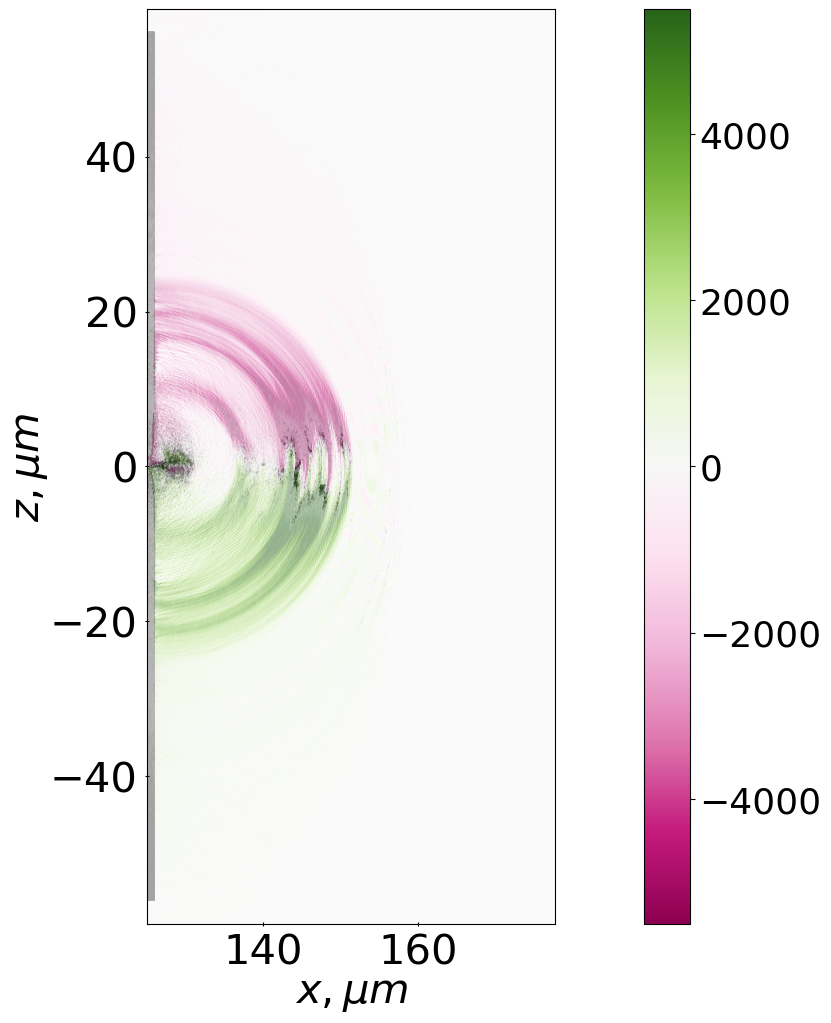} \hspace{0.1 \textwidth}
\includegraphics[width=0.4\textwidth]{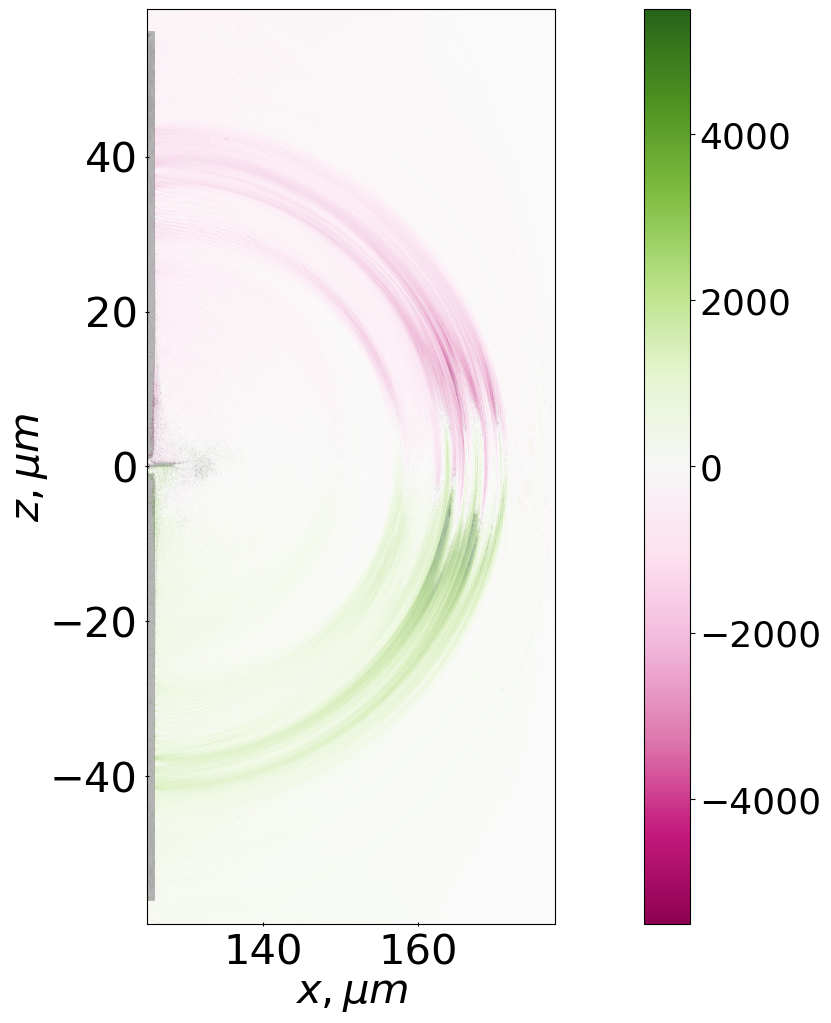}
\caption{The emitted electromagnetic field ($B_y$ component) behind the plasma-vacuum interface from the PIC simulation for 10 fs laser pulse at two instants. The legend is given in T.} \label{fig3}
\end{figure}

When accelerated electron beam crosses plasma-vacuum interface it generates transition radiation. Figure \ref{fig3} from the PIC simulation illustrates formation of a spherical electromagnetic wave propagating with the escaping electron bunch from the spot where the electron beam leaves a plasma.

In accordance with the estimates Eq. \eqref{epsmin} for the 30 fs laser pulse, only electrons with the energies in excess of 30 MeV contribute to transition radiation production. By using $\epsilon_{min}\simeq 30$ MeV one estimates the charge of the escaping electron beam, $Q_b\simeq 6.9$ nC, and its average energy, ${\cal E}_b \simeq 100$ MeV. The bunch of escaping electrons contains 97\% of energy of all accelerated electrons with the corresponding conversion efficiency of 31~\%.  Before escaping, the accelerated electrons occupy almost entire plasma cavity length, so that $L_b \simeq D \simeq \lambda_{p} \simeq 18 \lambda$. The electron bunch trasversal size is considerably less $D_b \sim 6-7 \lambda$. From these spacial scales the theoretical estimate Eq. (\ref{W}) gives $W_{THz} \simeq 0.13$J and $\sim 6$\% for the laser-to-THz conversion efficiency.
For the case of the 10 fs laser pulse with corresponding laser intensity and plasma density increase (see Eq. \eqref{rst}) the minimum required energy of escaping electrons grows to $\epsilon_{min} \simeq 50$ MeV without considerable effect on the escaping beam charge, $Q_b \simeq 7.5$ nC, what is the consequence of a well pronounced peak in the electron spectrum near the energy cut-off. Correspondingly, despite the fact that $\epsilon_{min}$ has increased
the capable of generating T-waves particle bunch, as before, contains 97~\% of all accelerated electrons. Using ${\cal E}_b \simeq 125$ MeV, $D_b \simeq 15 \lambda$, and $L_b \simeq 14 \lambda$ we obtain $W_{THz} \simeq 0.17$J with laser-to-THz conversion efficiency of 8\% from Eq. (\ref{W}). We found (see below) that Eqs. (\ref{W})) only slightly overestimate the radiation energy from simulations and can be advertised as being appropriate as easy-to-use estimates.

\begin{figure}[!ht]
\includegraphics[width=0.48 \textwidth]{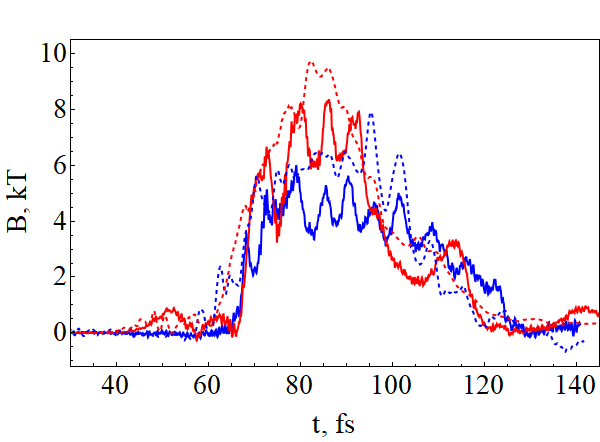}
\includegraphics[width=0.48 \textwidth]{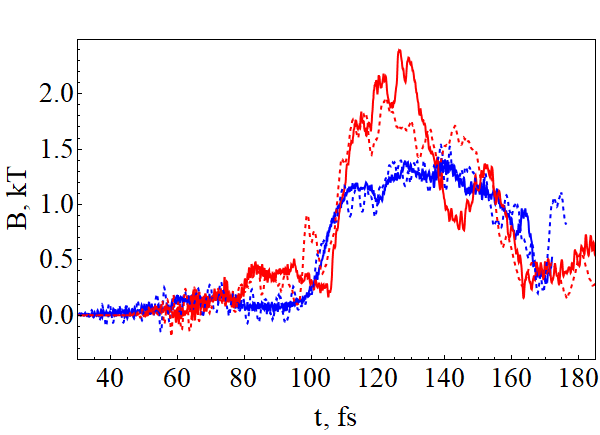}
\caption{Magnetic field component $B_y$ behind the target at the distance of $16\, \mu$m: (left) near the electron bunch, $z = 5 \,\mu$m and (right) closer to the periphery, $z = 25\, \mu$m, for the 10 fs (dashed red) and 30 fs (dashed blue) laser pulses.  The results for the case with additional foil are shown by the solid curves.}
\label{fig4}.
\end{figure}

The PIC simulation identified the propagating electromagnetic field in the short pulse form and the output from simulation was the evolution of all six field components at the given distance from the back side target surface. As example, Fig. \ref{fig4} shows $B_y$ magnetic field component near ($z = 5 \,\mu$m) and far ($z = 25 \,\mu$m) from the laser pulse propagation axis. In all cases  radiation field is practically in the form of half-cycle pulse (unipolar pulse). The difference between the cases with and without foil is rather small (cf. solid and dashed curves in Fig. \ref{fig4}).
The 1.5-2 times difference in the radiation field strength is clearly visible for the 10 fs and 30 fs laser pulses (cf. red and blue curves in Fig. \ref{fig4}) with the  intenser CTR for the shorter pulse.

Since accelerated electron beam experiences space-time modulation this manifests itself as corresponding modulation of the emitted electromagnetic pulse. Such modulation is well seen in Fig. \ref{fig4}. The measurement of the CTR modulation is able to reveal fine structure of the electron beam. This is what has already been proposed in Refs. \cite{vanTilborg06,glinec}.

\begin{figure}[!ht]
\includegraphics[width=0.45 \textwidth]{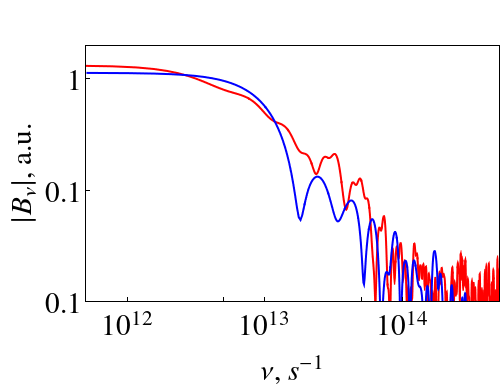}
\includegraphics[width=0.45 \textwidth]{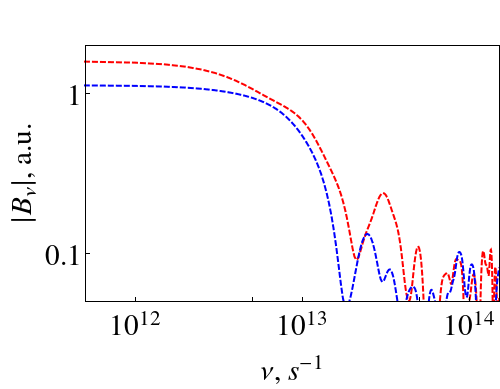}
\caption{Spectra of the CTR field at the distance of $z = 16\, \mu$m with deviation $z = 15\, \mu$m from the axis for the transparent plasma with an ultrathin foil at the target back side (left) and without foil (right side) for 10 fs laser pulse (red) and for 30 fs laser pulse (blue).}
\label{fig5}.
\end{figure}

The CTR field maximum is at small angle relative to electron bunch propagation direction, $\theta$, since ultrarelativistic electron energies. The theory predicts the maximum of the CTR intensity is at $\theta \sim 1/\gamma \simeq 0.3^\circ$ for considered electron bunch with average energy $\sim 100$ MeV.
The CTR electromagnetic field has a broadband spectrum with a cutoff at the characteristic frequency $\nu_c$ from several to ten THz. This spectrum is demonstrated in Fig. \ref{fig5}. The cuttoff is mostly defined by the electron beam length since $D_b$ is typically less than $L_b$. The estimate \cite{kuratov16,kuratov22}
$\nu_c = \omega_c/(2 \pi) \simeq c/(2 \pi L_b)$ fits the simulation results, $\nu_c \simeq 5$ THz ($L_b=15\mu$m) and $\nu_c \simeq 3$ THz ($L_b=10\mu$m).

\begin{figure}[!ht]
\includegraphics[width=0.85 \textwidth]{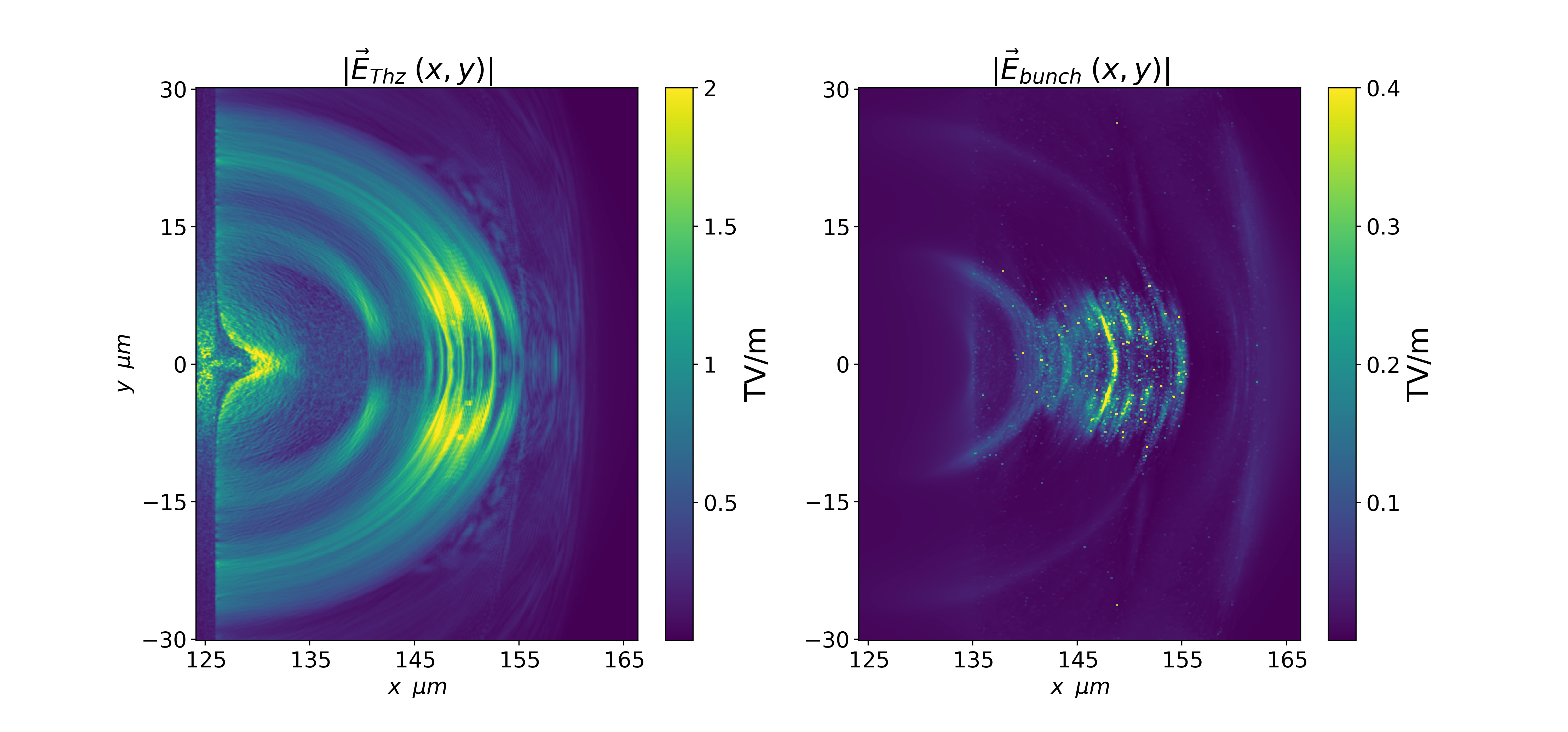}
\caption{Radiated (left panel) and intrinsic (right panel) electric fields of electron bunch for the case of 10 fs laser pulse with additional metal foil.}
\label{fig56}.
\end{figure}

\begin{figure}[!ht]
\includegraphics[width=0.6 \textwidth]{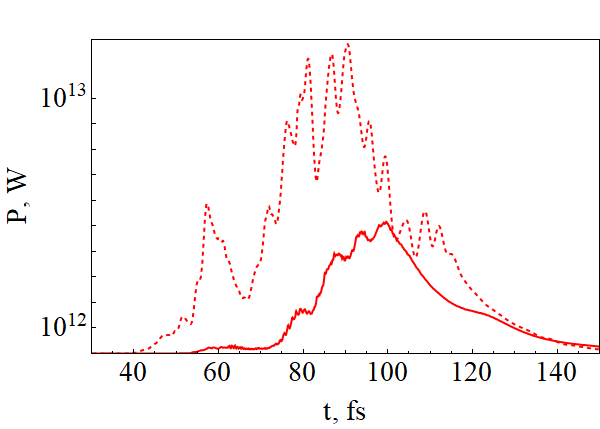}
\caption{Evolution of the radiated power at the surface with radius of 16 $\mu$m from target for the 10 fs laser pulse in the case with a foil at the back side (solid curve) and without a foil (dashed curve).}
\label{fig6}.
\end{figure}

The total electromagnetic field measured in the PIC simulation at the distance exceeding an electron beam length contains both the radiation field and the intrinsic (Coulomb) field of the electron bunch \cite{kuratov22}. We have developed an algorithm to find the contribution of only radiation field from the total registered electromagnetic emission by removing the electron bunch intrinsic field. The latter is given by the summation of the intrinsic fields of the macroparticles from the PIC distribution. The each macroparticle field is found through the Lienard-Wiechert potential \cite{Jackson}.
After that, we can restore only the radiated part of the electromagnetic field, i.e. estimate the CTR.

For the 10 fs laser pulse we get $W_{THz} \approx $0.16 J, which is approximately in a line with the expected analytical estimate, Eq. (\ref{W}). Comparison of the emitted and intrinsic fields is demonstrated by Fig \ref{fig56}. It is clearly seen that the Coulomb field of the electron bunch is maximum near bunch edges and disappears sharply away from them. The total energy of the bunch Coulomb field is about an order of magnitude less than the emitted energy, $W_C \simeq 0.02$ J, that roughly agrees with simple estimates for the field energy of the uniformly charged spherical-like electron cloud, $W_C \simeq Q_b^2/(D_b) \sim 0.03 $ J, with diameter $\sim D_b$. Note that bunch expands during propagation due to the transversal spread of electron velocities and Coulomb explosion, so that permanent decrease of the intrinsic field occurs.

Evolution of the radiated power was recorded at the spherical surface at the radius of 16 $\mu$m from the electron beam exit to vacuum. The obtained power pulse is shown in Fig. \ref{fig6}. For the case without a foil (dashed curve) the registered power contains dominant contribution of the transmitted laser light, while the metal foil reflects laser radiation and recorded power contains only CTR. In the latter case we may estimate the total radiation energy as $W_{THz} \simeq $0.12 J.

\section{Comparison with the THz CTR from a thin overdense target}\label{sec5}

To highlight the advantage of the RST regime of electron acceleration, we also simulated the interaction of the same laser pulse (2.2 J, 30 fs) with an Al target having a preplasma on the irradiated side. In the simulations, we used sharp density profile to model high-contrast laser pulse and the preplasma density profile found from hydrodynamic simulations \cite{rakitina} (see Sec. \ref{sec3}).

\begin{figure}[!ht]
\includegraphics[width=0.45 \textwidth]{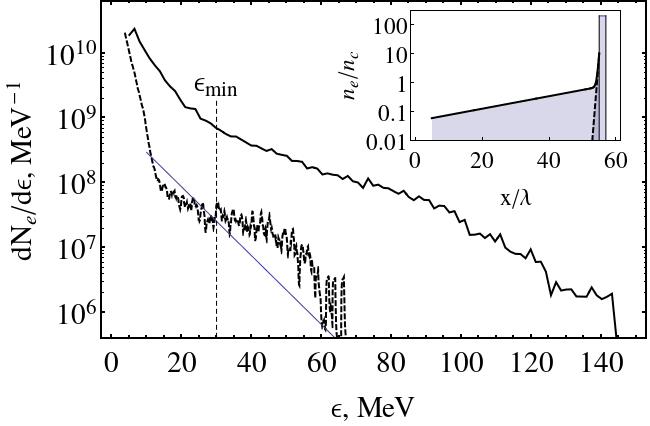}
\includegraphics[width=0.45 \textwidth]{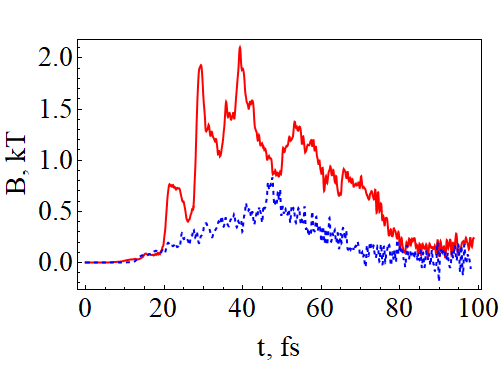}
\caption{Electron spectra generated by the 30 fs laser pulse in a target with extended preplasma  (solid curve) and short scale length preplasma (dashed curve) are shown in left panel. The used target density profiles are shown in the inset.  Thin blue line corresponds to the exponential spectrum with the ponderomotive temperature \cite{Wilks}.  Magnetic field component $B_y$ (in kT) behind the target at the distance of $10\, \mu$m and $z = 5 \,\mu$m for the target with extended preplasma (red solid line) and short scale length preplasma (blue dahsed line).}  \label{fig7}
\end{figure}

As expected \cite{Sentoku}, sufficiently extended preplasma, which can be produced by the ns ASE prepulse with an intensity of the order of $10^{13}$ W/cm$^2$ \cite{rakitina} significantly increases the efficiency of electron acceleration. Figure \ref{fig7} (left) shows comparison of the electron spectra from overdense plasma target with short and extended preplasma (see insert), which clearly demonstrates strong enrichment of the spectrum with high-energy electrons ($\epsilon > \epsilon_{min}$).

A presence of long preplasma is manifested in good acceleration of electron bunch, which total bunch finally reaches
$\sim 2.3$ nC accumulated in the particles with the energies in excess of 30 MeV. This corresponds to $\sim 5$ \% conversion efficiency into high-energy electrons. The maximum electron energy is in excess of hundred MeV. By using geometrical parameters of accelerated electron bunch, $L_b \simeq 20 \mu$m and $D_b \simeq 10 \mu$m,  for $\gamma_b \simeq 100$  we estimate a Thz energy from Eq. \eqref{W} as $W \simeq 12$ mJ, that is considerably higher than for the short preplasma case. However, this is significantly less than that obtained above for the electrons generated in a transparent plasma by laser pulse propagating in the RST regime (cf. Fig. \ref{fig4}). In Fig. \ref{fig7} (right) the latter reveals itself in about 3-times less magnetic field for the short preplasma as compared to the long one.

\section{Summary}\label{sec6}

Very efficient source of laser-triggered high-energy electrons generated in the regime RST of pulse propagation \cite{lobok18,bychenkov19} has been used for a study of the THz pulse production through coherent transition radiation. Similar to betatron radiation during electron acceleration in the RST regime \cite{Vais}, we have proven advantage of the considered THz source over the THz sources discussed earlier.

Although the analytical estimate, Eq. (\ref{W}) based only on the prediction of the RST electron acceleration \cite{lobok18,bychenkov19} demonstrates such advantage, we supplemented this estimate with quantitative confirmatory 3D PIC simulations. They are addressed to the multi-TW lasers which are currently not at all extraordinary and can work at about 10 Hz repetition rate. Our results show that the optimized laser-plasma parameters may provide up to 7\% conversion efficiency to super intense half-cycle THz pulses, that motivate targeted experiments on electron acceleration in the optimal RST regime and measurements of THz radiation. For PW laser installations, the laser-plasma interaction considered is capable to bring to generation of the super-high-energy THz pulses of the sub-joule level.

\section{Acknowledgments}

The authors acknowledge support from the Ministry of Science and Higher Education of the Russian Federation (agreement No. 075-15-2021-1361). A.V.B. acknowledges Russian Science Foundation (project No 24-22-00119) for support of the part with electron acceleration from the preplasma accompanying foils.

\section{Data availability }
The data that support the findings of this study are available from the corresponding author upon reasonable request.

\end{document}